\documentclass[
  aip,
  floatfix,
  jcp,
  twoside,
  twocolumn,
  10pt,
  showpacs,
  citeautoscript,
  superscriptaddress,
]{revtex4-1}

\usepackage{amsmath}
\usepackage{amssymb}
\usepackage{booktabs}
\usepackage{braket}
\usepackage{float}
\usepackage{glossaries}
\usepackage{graphicx}
\usepackage{tabularx}
\usepackage{units}
\usepackage{physics}
\usepackage{siunitx}
\usepackage{xspace}
\usepackage[version=4]{mhchem}
\usepackage[dvipsnames]{xcolor}
\usepackage[colorlinks=true,
            linkcolor=NavyBlue,
            citecolor=NavyBlue,
            filecolor=NavyBlue,
            urlcolor=NavyBlue
            ]{hyperref}

\DeclareSIUnit{\angstrom}{\textup{\AA}}

\renewcommand{\vec}[1]{\ensuremath\boldsymbol{#1}}

\newcommand{\gpaw}{GPAW}

\newcommand{\deltakick}{\texorpdfstring{\ensuremath{\delta}}{delta}-kick}

\setacronymstyle{long-short}
\newacronym{dc}{DC}{dipolar coupling}
\newacronym{dft}{DFT}{density functional theory}
\newacronym[plural="densities of state"]{dos}{DOS}{density of states}
\newacronym{dzp}{dzp}{double-$\zeta$ polarized}
\newacronym{dhct}{DHCT}{direct hot-carrier transfer}
\newacronym{fcc}{fcc}{face-centered cubic}
\newacronym{gllbsc}{GLLB-sc}{Gritsenko-van Leeuwen-van Lenthe-Baerends-solid-correlation}
\newacronym{hc}{HC}{hot carrier}
\newacronym{he}{HE}{hot electron}
\newacronym{homo}{HOMO}{highest occupied molecular orbital}
\newacronym{ks}{KS}{Kohn-Sham}
\newacronym{ksdft}{KS-DFT}{Kohn-Sham DFT}
\newacronym{lcao}{LCAO}{linear combination of atomic orbitals}
\newacronym{ld}{LD}{Landau damping}
\newacronym{lsp}{LSP}{localized surface plasmon}
\newacronym{lrtddft}{LR-TDDFT}{linear-response TDDFT}
\newacronym{lumo}{LUMO}{lowest unoccupied molecular orbital}
\newacronym{np}{NP}{nanoparticle}
\newacronym{paw}{PAW}{projector augmented wave}
\newacronym{pdos}{PDOS}{projected DOS}
\newacronym{rto}{RTO}{regular truncated octahedron}
\newacronym{rttddft}{RT-TDDFT}{real-time time-dependent density functional theory}
\newacronym{sc}{SC}{strong coupling}
\newacronym{si}{SI}{Supplementary Information}
\newacronym{tddft}{TDDFT}{time-dependent density functional theory}
\newacronym{xc}{XC}{exchange correlation}

\newcommand{\phys}{
    Department of Physics,
    Chalmers University of Technology,
    SE-412~96 Gothenburg, Sweden
}
\newcommand{\aalto}{
    Department of Applied Physics,
    Aalto University,
    00076 Aalto, Finland
}

\begin{document}

\title{
    Hot-carrier transfer across a nanoparticle-molecule junction: \texorpdfstring{\\}{ }
    The importance of orbital hybridization and level alignment
}

\author{Jakub\ Fojt}
\affiliation{\phys}

\author{Tuomas\ P.\ Rossi}
\affiliation{\aalto}

\author{Paul\ Erhart}
\email[Corresponding author: ]{erhart@chalmers.se}
\affiliation{\phys}

\begin{abstract}
While direct hot-carrier transfer can increase photo-catalytic activity, it is difficult to discern experimentally and competes with several other mechanisms.
To shed light on these aspects, here, we model from first principles hot-carrier generation across the interface between plasmonic nanoparticles and a CO molecule.
The hot-electron transfer probability depends non-monotonically on the nanoparticle-molecule distance and can be effective at long distances, well outside the region of chemisorption; hot-hole transfer on the other hand is limited to shorter distances.
These observations can be explained by the energetic alignment between molecular and nanoparticle states as well as the excitation frequency.
The hybridization of the molecular orbitals is the key predictor for hot-carrier transfer in these systems, emphasizing the need to include the effects of ground state hybridization for accurate predictions.
Finally, we show a non-trivial dependence of the hot-carrier distribution on the excitation energy, which could be exploited when optimizing photo-catalytic systems.
\end{abstract}

\maketitle

\section{Introduction}

Plasmonic metal \glspl{np} are fundamental components in several emerging technologies, including sensing \cite{NugDarCus19, DarKhaTom21}, light-harvesting \cite{GenAbdBer21}, solar-to-chemical energy conversion \cite{AslRaoCha18, LiCheRic21, DuCTagWel18} and catalysis \cite{ZhoLouBao21, DuCTagWel20, HouCheXin20, YamKuwMor21}.
The properties that set these materials apart for these applications are their high surface-to-volume ratios and high optical absorption cross sections at visible frequencies \cite{Boh83, LanKasZor07}, the latter being due to the presence of a \gls{lsp} resonance \cite{KreVol95}.
In particular plasmonically-driven catalysis is an active research field, addressing important chemical reactions such as ethylene epoxidation, CO oxidation or \ce{NH3} oxidation that are catalyzed by illuminating \glspl{np}, e.g., of the noble metals Ag \cite{ChrXinLin11, YamKuwMor21}, Au \cite{DuCTagWel18, LiCheRic21, SahYanMas22} or Cu \cite{DuCTagWel20, HouCheXin20}.

The \gls{lsp}, which is a collective electronic excitation, is excited by absorption of light and decays within tens of femtoseconds \cite{BerMusNea15, RosKuiPus17, ZhoSweZha18, AslRaoCha18, RosErhKui20, KumRosKui19, KumRosMar19, RosErhKui20, VilLeiMar22} into a highly non-thermal (usually referred to as ``hot'') distribution of electrons and holes \cite{BroHalNor15, GonMun15, RomHesLis19, Khu19, Khu20, HatMenZhe21, HawSilBer21}.
Chemical reactions can then be catalyzed by \glspl{hc} transiently populating orbitals of nearby molecules \cite{LinAslBoe15, AslRaoCha18}, which can, e.g., lower reaction barriers \cite{ZhoSweZha18}.
Two variants of this process can be distinguished.
In the \textit{direct} \gls{hc} transfer process \cite{KhuPetEic21} (also known as chemical interface damping) the \gls{lsp} decays into an electron-hole pair, where one of the carriers is localized on the reactant molecule and the other on the \gls{np}.
In the \textit{indirect} \gls{hc} transfer process both carriers are generated in the \gls{np}, and at a later time scattered into molecular orbitals.
The efficiency and importance of these processes as well as their competition with thermal effects, is still a matter of intense debate \cite{DubUnSiv20, Jai20, ZhoSweZha18, SivBarUn19, ZhoSweRob19}, highlighting the importance of more detailed atomistic studies.
The direct \gls{hc} transfer process is promising in terms of efficiency and selectivity \cite{ChrXinLin11, LinAslBoe15}, and has been studied experimentally \cite{SeeTheLou19}, in theory \cite{KhuPetEic21}, and by computational \textit{ab-initio} models \cite{MaGao19, KumRosMar19, KumRosKui19}.
Typically the focus lies on understanding \gls{hc} generation at surfaces \cite{SunNarJer14, SeeTheLou19}, but there has not yet been a detailed account of the dependence of \gls{hc} transfer on molecular position and orientation, and whether there are handles for tuning \gls{hc} devices to particular molecules in all probable states of thermal motion.
Yet these aspects are crucial for direct \gls{hc} transfer processes, which exhibit an intricate dependence on the hybridization of molecular and surface states as also shown in this work.

In this work, we study plasmon decay and carrier generation across a \gls{np}-molecule junction, which is the initial step in the direct \gls{hc} transfer process.
We consider plasmonic Ag, Au, and Cu \glspl{np} in combination with a CO molecule, whose excitations are energetically much higher than the plasmon resonance of any of the \glspl{np} considered here.
In a \gls{rttddft} \cite{YabBer96} framework, we drive the system with an ultra-fast laser pulse to induce a plasmon.
We simulate the electron dynamics in the system until the plasmon has decayed, and then analyze the distribution of carriers over the ground state \gls{ks} states.
To this end, we employ and extend the analysis methods previously developed by us \cite{KumRosKui19, KumRosMar19, RosErhKui20}.

We consider a wide range of geometrical configurations.
Specifically, we treat (111) on-top, (111) \gls{fcc}, (100) hollow and corner sites of molecular adsorption on the \gls{np} and various distances between the molecule and \gls{np}.
We map out the \gls{hc} transfer efficiency as a function of the \gls{np}-molecule geometry (adsorption site, distance, molecular bond length), excitation energy, and material (Ag, Au, Cu).

The article is structured as follows.
We present and discuss our results for Ag \gls{np}+CO systems driven by a laser pulse tuned to the \gls{lsp} resonance in \autoref{subsec:geometry-dependence}, \autoref{subsec:level-alignment} and \autoref{subsec:site-distribution}.
In \autoref{subsec:pulse-dependence} we quantify the dependence of carrier generation on the pulse frequency, including Ag, Au, and Cu \glspl{np}.
Finally, in \autoref{sec:conclusions} we discuss the implications of our findings.
Details concerning our methodology, the systems under study as well as the parameters used in computations are provided in \autoref{sec:methods}.

\section{Results and discussion}
\label{sec:results}

\subsection{Adsorption geometry-dependent carrier generation in Ag}
\label{subsec:geometry-dependence}

\begin{figure*}
    \centering
    \includegraphics{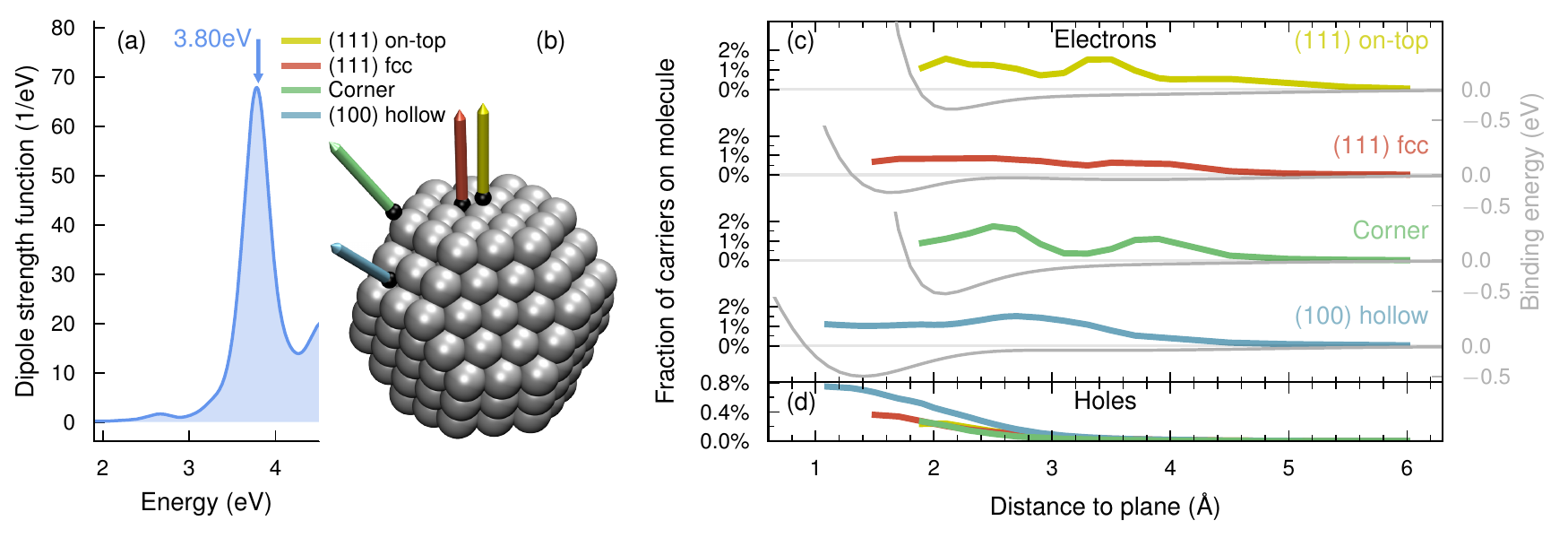}
    \caption{
        Geometry dependence of \gls{hc} generation in \ce{Ag201} \gls{np}+CO.
        (a) Optical spectrum of the bare \gls{np}.
        The frequency \SI{3.8}{\electronvolt} of the driving laser is marked by an arrow above the spectrum.
        (b) Model of the \gls{np} with the axes along which the \gls{np}-molecule distance is varied.
        (c-d) Fractions of generated electrons (c) and holes (d) on the molecule (\autoref{eq:region-electron-generation}) after \SI{30}{\femto\second} and binding energies (\autoref{eq:binding-energy}) as a function of distance and site.
    }
    \label{fig:geometry-dependence}
\end{figure*}

We consider the CO molecule at a range of distances from the \ce{Ag201} \gls{np} (111) on-top, (111) \gls{fcc}, (100) hollow and corner sites.
The plasmon (\SI{3.8}{\electronvolt}, \autoref{fig:geometry-dependence}a) and first optical excitation of CO (\SI{14.5}{\electronvolt}, \autoref{sfig:CO}) are not resonant and the optical response of the combined \gls{np}+CO system is not strongly dependent on geometry (\autoref{sfig:lspr}).

The bond length of CO, which is \SI{1.144}{\angstrom} in the free molecule, increases when adsorbed to the \gls{np} (corner: \SI{1.150}{\angstrom}, (111) on-top: \SI{1.151}{\angstrom}, (111) \gls{fcc}: \SI{1.170}{\angstrom} and (100) hollow \SI{1.184}{\angstrom}).
Higher coordination numbers for the C atom thus result in larger bond lengths.
This is in agreement with previous studies on the extended (111) Ag surface \cite{GajEicHaf04}, and can be understood by considering that as C shares more electron density in bonds with metal atoms, the CO bond is weakened.

As we vary the \gls{np}-CO distance (\autoref{fig:geometry-dependence}b), we observe minima in the binding energy curves around \SI{250}{\milli\electronvolt}-\SI{500}{\milli\electronvolt} at \SI{1.5}{\angstrom}-\SI{2.3}{\angstrom} (\autoref{fig:geometry-dependence}c).
Here, we define the binding energy as
\begin{align}
    \label{eq:binding-energy}
    E^\text{(site)}_\text{bind}(d) = E^\text{(site)}(d) - E_\text{NP}^\text{(site)} - E_\text{mol}^\text{(site)},
\end{align}
where $E_\text{NP}^\text{(site)}$ and $E_\text{mol}^\text{(site)}$ are the energies of the \gls{np} and molecule, respectively, taken from two separate calculations, representing infinite separation.
Note that in this definition (\autoref{eq:binding-energy}), we take $E_\text{NP}^\text{(site)}$ and $E_\text{mol}^\text{(site)}$ as the energies of the \gls{np} and molecule as in the relaxed configuration for the specific site, emphasized by the superscript (site).
Allowing the molecule to relax at each distance effectively widens the adsorption curve but does not affect the main conclusions drawn here (\autoref{sfig:binding_relax}, \autoref{sfig:hc-generation}).
Fixing the bond length reduces, however, the degrees of freedom and simplifies the following discussion, whence we adopt this constraint here.

We drive the Ag \gls{np}+CO system with a Gaussian laser pulse tuned to the \gls{lsp} frequency $\hbar \omega = \SI{3.8}{\electronvolt}$.
Within the first tens of femtoseconds a plasmon forms in the \gls{np} and decays into resonant excitations, for which the electron-hole energy difference equals $\hbar\omega$.
The plasmon formation and decay process in similar systems has previously been studied in detail by our group \cite{RosKuiPus17, RosErhKui20} and is not covered here.

We then measure the fraction of generated electrons in the molecule (\autoref{eq:region-electron-generation}) after plasmon decay (\autoref{fig:geometry-dependence}c).
While intuition would suggest this quantity to decrease monotonically with decreasing wave function overlap at increasing distances, we find the fraction of generated \glspl{hc} to be of similar magnitude measuring between 0.5 and \SI{2}{\%} over a wide range of distances with several site specific features.
A smooth decay to zero only occurs beyond 4 to \SI{5}{\angstrom}.
Below this threshold several of the sites feature one or two peaks, including near \SI{2.1}{\angstrom} and \SI{3.3}{\angstrom} for the (111) on-top site, \SI{2.7}{\angstrom} for the (100) hollow site and \SI{2.7}{\angstrom} and \SI{3.9}{\angstrom} for the corner site.
Only the (111) \gls{fcc} site appears relatively feature-less.

By contrast, the binding energies depend smoothly on distance and approach zero already at 3 to \SI{4}{\angstrom}.
The landscape of electron generation on the molecule thus extends further than the features in the potential energy surface and is more sensitive to the underlying shifts in eigenenergies and wave function overlaps.
Our findings imply that across-interface \gls{hc} generation can be effective even at quite long distances (up to \SI{5}{\angstrom}) from the \gls{np}, and does not require molecular adsorption.

The fraction of holes generated on the molecule (\autoref{eq:region-hole-generation}), on the other hand, decays smoothly with distance (\autoref{fig:geometry-dependence}d) reaching a maximum of \SI{0.2}{\%} to \SI{0.8}{\%}.

\subsection{Origin of the \texorpdfstring{\gls{hc}}{HC} generation distance dependence}
\label{subsec:level-alignment}

\begin{figure}
    \centering
    \includegraphics{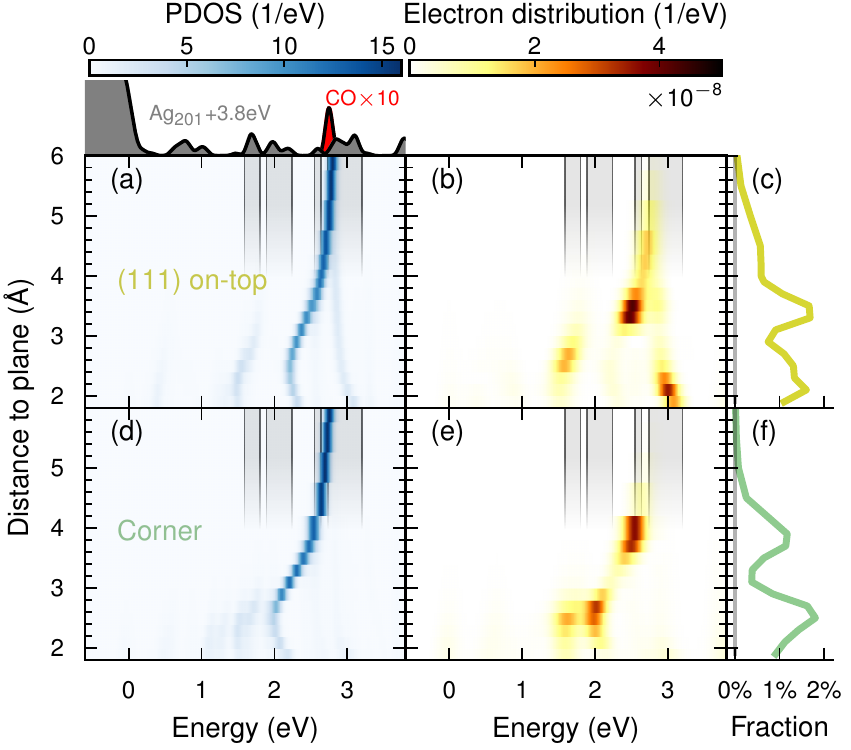}
    \caption{
        Level alignment between molecular and \gls{np} \gls{pdos} for (111) on-top (a-c) and corner (d-f) sites.
        (a, d) Molecular \gls{pdos} as a function of distance.
        As the molecule approaches the \gls{np} the \gls{lumo} shifts to lower energies, eventually splitting into several branches.
        The \gls{pdos} for the \gls{np} and molecule at far separation are indicated above the plot, where the \gls{np} \gls{pdos} has been shifted by the pulse frequency.
        Shaded regions correspond to (a selection of) large values in the shifted \gls{np} \gls{pdos}.
        (b, e) Electron distribution as a function of distance. (c, f) The fraction of electrons generated in the molecule.
    }
    \label{fig:level-alignment}
\end{figure}

To explain the rich behavior of the across-interface electron generation as a function of distance we study the molecular \gls{pdos}  (\autoref{fig:level-alignment}a,c) and the energy distribution of \glspl{hc} generated on the molecule (\autoref{eq:region-energetic-electron-generation}, \autoref{fig:level-alignment}b,d).
We note that as the molecule only contains a small fraction of the electrons in the system, the \gls{np} \gls{pdos} makes up the most part of the total \gls{dos} and is practically independent of distance.
The molecular \gls{pdos}, however, is strongly site and distance-dependent.
At long distances the \gls{pdos} is comprised of a single \gls{lumo} level at \SI{2.8}{\electronvolt}, which shifts to lower energies with decreasing distance, eventually splitting into several branches.
Most of these branches are above the Fermi level (\autoref{sfig:pdos}) and thus represent unoccupied hybridized states.

The hot electron distribution in the molecule clearly mirrors the shape of the \gls{pdos} as electrons are generated in unoccupied molecular levels.
The distribution of electrons in each \gls{pdos} branch is, however, not simply proportional to the corresponding \gls{pdos} weight.
As the transitions ($\varepsilon_i \rightarrow \varepsilon_a$) induced by the pulse are resonant with the pulse frequency ($\varepsilon_a - \varepsilon_i = \hbar \omega_\text{pulse} = \SI{3.8}{} \pm \SI{0.37}{\electronvolt}$, the value after $\pm$ denotes the half-width at half-maximum of the Gaussian pulse in frequency space), the electron distribution is determined by the energetic alignment of the molecular \gls{pdos} with the \gls{np} \gls{pdos}.
It is worth noting that various states in the \gls{np} couple with differing strength to the molecular states; the alignment with particular peaks in the \gls{np} \gls{pdos} is therefore important (\autoref{sfig:hcdist-map})
The across-interface electron generation is thus enhanced at energies $\varepsilon$ where (1) the molecular \gls{pdos} is large at $\varepsilon$, (2) the \gls{np} \gls{dos} is large at $\varepsilon - \hbar\omega$ and (3) the transition dipole moment between the corresponding \gls{np} and molecular states is sizable.
It is important to emphasize that it is the combination of these effects that dictates the response.
It is, however, usually much simpler to obtain the \glspl{pdos} than the transition dipole matrix elements, and we may assess the basic possibility for \gls{hc} transfer already on this basis.
This is apparent, e.g., in the decomposition of the electrons distribution in terms of the underlying single-particle excitations (\autoref{sfig:hcdist-map}).
The variation of the \gls{hc} transfer probability with distance is the result of transitions from two occupied states \SI{0.95}{} and \SI{1.20}{\electronvolt} below the Fermi energy and one unoccupied state \SI{2.46}{\electronvolt}.
The transitions involving these three states at \SI{3.41}{} and \SI{3.66}{\electronvolt} are slightly different from the excitation frequency of \SI{3.8}{\electronvolt} considered in this section but are still activated due to the finite linewidth of the excitation pulse.
In fact, in \autoref{subsec:pulse-dependence} below, we will find that a slight reduction in the excitation frequency leads to a notable increase in the \gls{hc} transfer probability.

In a similar manner we can understand the across-interface generation of holes, with the rule that an occupied state $\varepsilon$ in the molecule must align with a peak in the \gls{np} \gls{pdos} at $\varepsilon + \hbar\omega$.
As the CO \gls{homo} level is at \SI{-4.8}{\electronvolt} in the free molecule (long distance limit), hole generation is not possible with the pulse frequency \SI{3.8}{\electronvolt}.
Transfer is only possible at close distances where hybridized branches of the \gls{homo} and \gls{lumo} appear in the region $-\SI{3.8}{\electronvolt} < \varepsilon < \SI{0}{\electronvolt}$, beginning at distances around \SI{2.5}{\angstrom} (\autoref{sfig:pdos}).

The energetic level alignment is thus a good descriptor in predicting across-interface \gls{hc} generation.
Other factors, such as the amount of wave function overlap and the orbital momentum character of states, play a role, i.e., in determining the coupling strength between occupied and unoccupied states, but the energetic level alignment is sufficient to qualitatively explain the distance and site dependence.
In the case of \ce{Ag201} with \SI{3.8}{\electronvolt} pulse frequency the constructive contributions to the \gls{lsp} stem from delocalized sp-band states \cite{RosKuiPus17} that also have a larger spatial extent.
This provides a rationale for the rather long-ranged effect that we observe here.
Based on this insight, one could expect the effect to be shorter ranged in non-noble metals such as Pd and Pt, for which the Fermi level lies inside the d-band.
A deeper investigation of this question is, however, beyond the scope of the present study.
Finally, we note that for larger \glspl{np}, where the \gls{dos} between d-band onset and Fermi level is more smeared out, we expect the energetic level alignment to be less noticeable.

\subsection{Comparison of across-interface electron generation to surface electron distribution}
\label{subsec:site-distribution}

\begin{figure}
    \centering
    \includegraphics{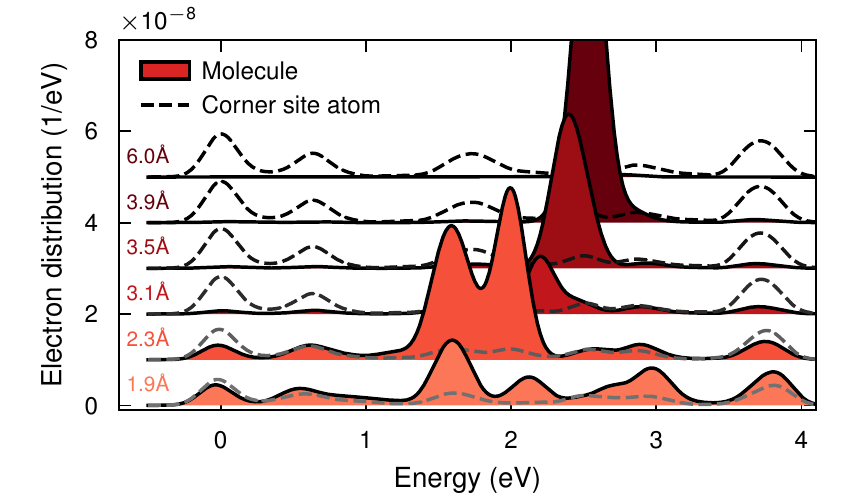}
    \caption{
        Energy distribution of electrons generated on the molecule and on the \gls{np} corner site that the molecule approaches.
        While the former varies non-monotonically, the latter is practically unchanged with distance.
    }
    \label{fig:site-distribution}
\end{figure}

To further emphasize that the energy distribution of electrons generated on the molecule depends on energetic level alignment, we compute the energy distribution of electrons on the nearest metal atom (the integral of \autoref{eq:region-energetic-electron-generation} in the metal-atom Voronoi cell) as a function of distance (\autoref{fig:site-distribution}).
We focus specifically on the corner site, where the distance dependence of the electron distribution on the molecule exhibits two clear maxima.
We find that the electron distribution on the adsorption site is practically distance-independent.

Only at the smallest considered distances does the electron distribution on the molecule resemble the electron distribution at the adsorption site on the metal.
Hence it is not enough to know the electron distribution on surfaces and surface sites for a bare \gls{np} to predict across-interface electron generation in combined \gls{np} + molecule systems.
Equipped with distributions for the bare \gls{np} alone one misses for example that about 2 times more electrons are generated at \SI{2.5}{\angstrom} for the corner site, than at \SI{1.9}{\angstrom}.
In other words, the surface electron distribution is an insufficient predictor for across-interface electron generation.

\subsection{Adsorption geometry and pulse frequency-dependent \texorpdfstring{\gls{hc}}{HC} generation}
\label{subsec:pulse-dependence}

\begin{figure*}
    \centering
    \includegraphics{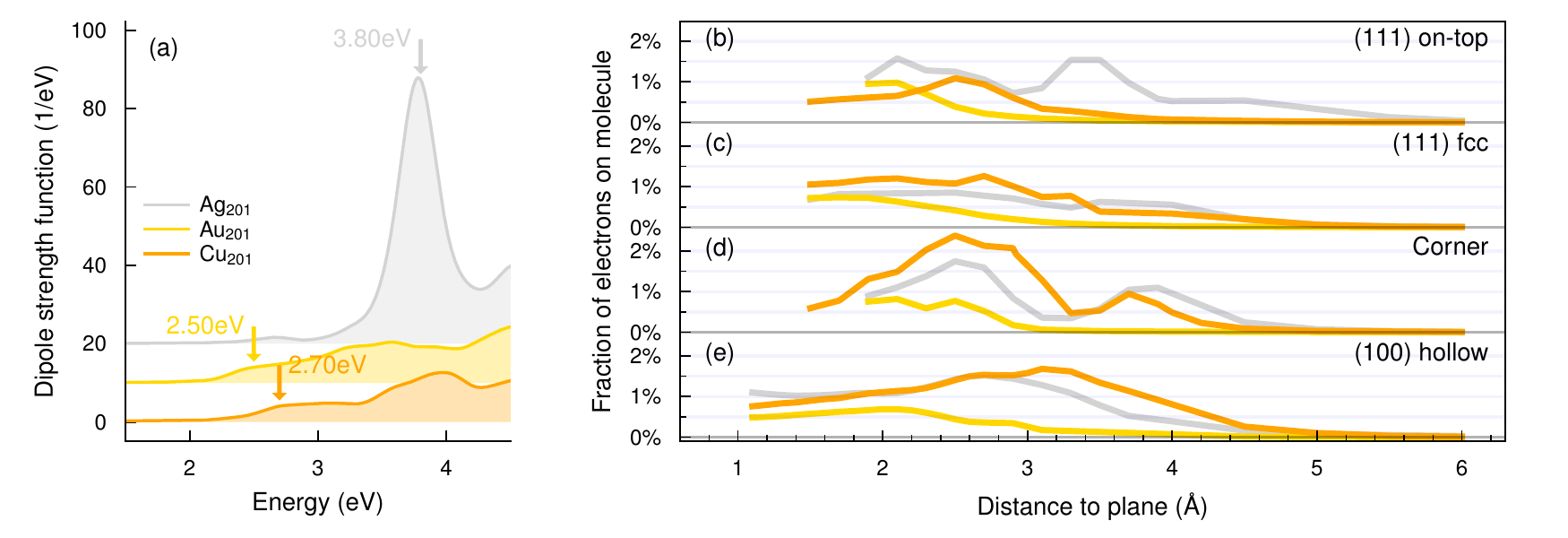}
    \caption{
        Geometry dependence of electron generation in \ce{Ag201}, \ce{Au201} and \ce{Cu201} \glspl{np}+CO.
        (a) Optical spectra of the bare \glspl{np}.
        (b-e) Fractions of generated electrons on the molecule (\autoref{eq:region-electron-generation}) after plasmon decay as a function of distance and site with pulse frequency \SI{3.8}{\electronvolt} (Ag), \SI{2.5}{\electronvolt} (Au) and \SI{2.7}{\electronvolt} (Cu).
    }
    \label{fig:element-dependence}
\end{figure*}

\begin{figure*}
    \centering
    \includegraphics{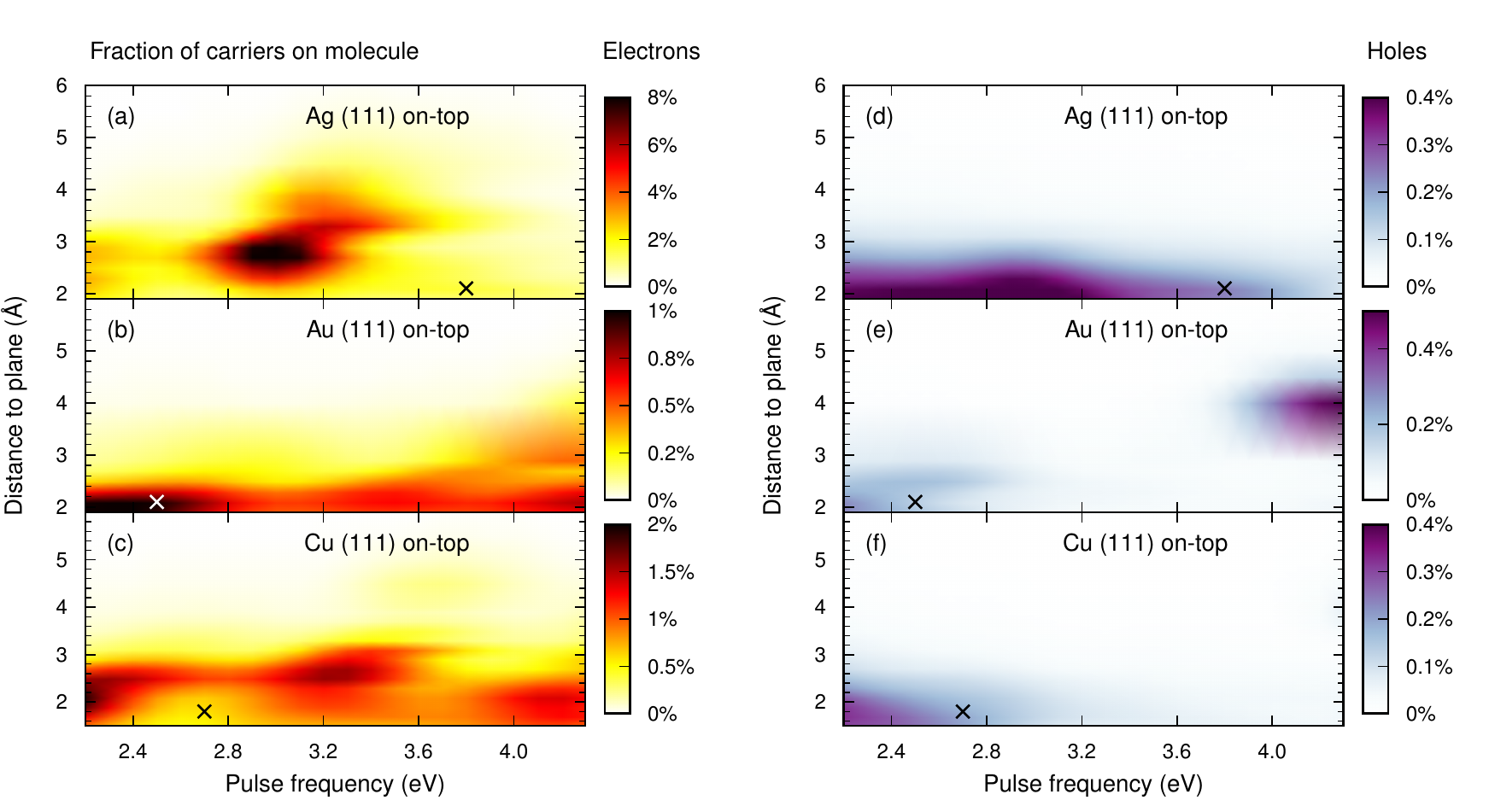}
    \caption{
        Pulse-frequency dependence of the electron generation in CO.
        Fraction of generated electrons (a-c) and holes (d-f) on the molecule as a function of distance and pulse frequency for the (111) on-top site in Ag (a, d), Au (b, e) and Cu (c, f).
        For reference, the crosses in the figure mark the distance corresponding to the adsorption minimum, and the pulse frequency used in \autoref{fig:element-dependence}.
    }
    \label{fig:pulse-dependence}
\end{figure*}

We now extend our study to also include Au and Cu.
The s-electrons have nearly identical \glspl{dos} in the \ce{Ag201}, \ce{Au201} and \ce{Cu201} \glspl{np} but the d-band onsets differ (Ag: \SI{3.7}{\electronvolt}, Au: \SI{2.1}{\electronvolt}, Cu: \SI{2.3}{\electronvolt} below the Fermi level; \autoref{sfig:NP-dos}).
As a consequence of the earlier d-band onset \ce{Au201} and \ce{Cu201} lack the well defined \gls{lsp} peak of \ce{Ag201}\cite{CazDolRub00} (\autoref{fig:element-dependence}a).
The binding energy curves are also similar to Ag, with the main difference that the molecule binds more strongly and closer to Cu (\autoref{sfig:binding}).

We drive the \gls{np}+CO with a Gaussian laser pulse (Ag: frequency \SI{3.8}{\electronvolt}, Au: \SI{2.5}{\electronvolt}, Cu: \SI{2.7}{\electronvolt}) and measure the fraction of electrons generated on the molecule (\autoref{fig:element-dependence}b-e).
We observe similar trends in Ag and Cu, both exhibiting peaks near \SI{2.1}{\angstrom} for the (111) on-top site, \SI{2.7}{\angstrom} for the (100) hollow site and \SI{2.7}{\angstrom} and \SI{3.9}{\angstrom} for the corner site.
Only the \SI{3.3}{\angstrom} peak in the (111) on-top site of Ag lacks a counterpart in Cu.
In contrast to Ag and Cu, the Au \gls{np} shows smooth trends, without pronounced peaks, of decreasing electron generation on the molecule with increasing distance.

The similarity in electron generation for Ag and Cu can be explained by a similar distance dependence of the molecular orbital hybridization (\autoref{sfig:pdos}).
We note that while the resonance condition is not the same for Ag and Cu ($\hbar\omega=\SI{3.8}{\electronvolt}$ and \SI{2.7}{\electronvolt}, respectively), the similar energy-spacing between hybridized molecular levels is enough to yield similar electron generation curves.
The \SI{3.3}{\angstrom} peak is the only clear feature that is missing in Cu, the reason being that the molecular orbital is too far from the Fermi level (\SI{2.8}{\electronvolt}, to be compared to $\hbar\omega=\SI{2.7}{\electronvolt}$).
The hybridization behavior in Au differs from the behavior in Ag and Cu.
At long distances the CO \gls{lumo} is further from the Fermi level in Au than in Ag or Cu (due to Au having a higher work function, and us considering different bond length of the molecule for each metal), preventing electron generation.
As the distance decreases the orbital hybridizes more strongly, splitting into more branches.
As a consequence, at small distances there are more \gls{pdos} branches in which electron generation occurs, leading to a smoother distance dependence.

Based on our observations we should expect the pulse frequency $\hbar\omega$ to act as a handle for tuning the electron generation through the approximate (barring electron-hole coupling) resonance condition $\varepsilon_a - \varepsilon_i = \hbar\omega$.
Indeed, the electron generation depends non-monotonically on both distance and pulse frequency (\autoref{fig:pulse-dependence}a-c).
For example, by lowering the pulse frequency we can get rid of the dip in electron generation at \SI{2.9}{\angstrom} for the Ag (111) on-top site; using $\hbar\omega =\SI{3.1}{\electronvolt}$ the feature at the same distance instead becomes a maximum.
Choosing the pulse frequency appropriately the fraction of electrons generated on the molecule can be as high as \SI{8.9}{\%} for Ag (distance \SI{2.7}{\angstrom}, $\hbar\omega =\SI{3.1}{\electronvolt}$),  \SI{1.1}{\%} for Au (distance \SI{1.9}{\angstrom}, $\hbar\omega =\SI{2.2}{\electronvolt}$), and  \SI{2.3}{\%} for Cu (distance \SI{2.1}{\angstrom}, $\hbar\omega =\SI{2.2}{\electronvolt}$).
While the electron generation in the Ag and Cu systems shows a complex dependence on pulse frequency and distance, in the case of Au, it is almost monotonic in both dimensions.
However, in both endpoints of the considered pulse frequencies the behavior for Au becomes more interesting; at low frequencies the distance dependence becomes very sharp, as fewer \gls{pdos} branches fall into the relevant energy range, and at high frequencies there is electron generation at long distances, due to the free-molecule \gls{lumo} falling into the relevant energy range.

In the case of Ag (\autoref{fig:pulse-dependence}d) and Cu (\autoref{fig:pulse-dependence}f), holes are generated at small distances (where hybridized branches of the \gls{lumo} orbital are below the Fermi energy; \autoref{sfig:pdos}) with a weak dependence on pulse frequency.
For Au (\autoref{fig:pulse-dependence}e) hole generation becomes relatively strong at intermediate distances (\SI{3}{} - \SI{5}{\angstrom}) and large pulse frequencies.
This behavior originates from the \gls{homo} orbital, which in the long-distance limit resides \SI{3.9}{\electronvolt} below the Fermi energy, but shifts to lower energies with decreasing distance (i.e., out of the range $\varepsilon_a - \varepsilon_i = \hbar\omega, \varepsilon_a > 0$, thus limiting hole generation).

The frequency of the exciting light is thus an excellent handle for tuning the fraction of carriers generated on the molecule, which is especially interesting in applications where selectivity is important.
In molecules with several orbitals close enough to the Fermi energy to be optically accessible, the generation of electrons in one orbital could be favored over the other.
It is, however, important at this stage to remember that changing the pulse frequency also changes the total optical absorption and thus the total number of generated carriers.
We therefore also consider the total, pulse-frequency and distance-dependent, amount of electrons generated on the molecule (\autoref{sfig:pulse-dependence}, that is, contrary to before, \emph{without} expressing it as a fraction of electrons generated in the entire \gls{np}+CO system).
In particular for the Ag \gls{np}, which has an exceptionally sharp absorption spectrum, the pulse dependence is affected, with a maximum in total electron generation on the molecule occurring using a pulse frequency of \SI{3.6}{\electronvolt} (to be compared to a maximum in the fraction of electrons generated on the molecule at \SI{3.1}{\electronvolt}).
As a final note, we point out that it should be possible to simultaneously tune the pulse frequency to the desired resonance condition $\varepsilon_a - \varepsilon_i = \hbar\omega$, and the \gls{lsp} resonance of the \gls{np} (thus the optical absorption) to the pulse frequency, by taking advantage of the fact that the \gls{lsp} is more sensitive the size and shape of the \gls{np} than, e.g., the \gls{dos}.

\section{Conclusions and outlook}
\label{sec:conclusions}

In this study we have investigated the geometry dependence of \gls{hc} generation across noble metal-molecule interfaces due to plasmon absorption and decay.
We have found that typically up to 0.5 -- \SI{3}{\%} of all electrons generated in the system end up on the molecule after plasmon decay, even up to distances of \SI{5}{\angstrom}, which is well outside the region of chemisorption.
By tuning the excitation frequency, we are able to achieve up to \SI{8.9}{\%} electrons generated in the molecule, at the expense of lower absolute amount of electrons generated.
These findings suggest that direct \gls{hc} transfer is a relevant process in plasmon decay, and that the process does not require chemisorption of molecules to the absorbing medium.

We have also shown that the fraction of generated electrons on the molecule depends on the geometry of the molecule and \gls{np} in rather intricate fashion.
This geometry dependence can be understood in terms of the energy landscape of hybridized molecular orbitals; as an orbital shifts so it is resonant with certain peaks in the \gls{np} \gls{pdos} an increase in the probability for \gls{hc} transfer can be expected.
The distance-dependent behavior of the hybridization differs enough for the various sites so that also the carrier generation differs.
For larger \glspl{np}, these effects could be less important, as the \gls{dos} would be smoother.

In \gls{hc} transfer processes the rate of charge carrier generation across the metal-molecule interface is in competition with various loss channels, such as the rates of reemission, and scattering and subsequent thermalization of excited carriers with phonons, surfaces, and other carriers \cite{BerMusNea15, Khu19}.
As these loss channels are currently beyond the reach of our calculations, we have to view our results as an upper bound on the efficiency of \gls{hc} transfer, i.e., out of all photon absorption events that occur, up to 0.5 - \SI{3}{\%} (or \SI{8.9}{\%} when tuning the excitation frequency) result in electron transfer to the molecule.
It is worthwhile to point out that \gls{hc} generation is a quantized process, where it is very unlikely that there is at one time more than one excited plasmon at a time, under illumination conditions that are realistic for energy-harvesting applications \cite{Khu19, Khu20}.
Each plasmon decays into one electron-hole pair where one carrier can be either in the molecule or in the metal, and we should thus consider our computed fractions as probabilities.

While the \gls{hc} transfer landscape has a detailed structure, it is important to remember that in reality both the molecule and the \gls{np} are subjected to thermal motion.
Since it is the weakest interaction in the combined system, one can expect the relative motion of \gls{np} and molecule to have the most pronounced thermal effect on level alignment, effectively broadening the peaks in the \gls{hc} transfer probability curves.
Considering $kT \approx \SI{25}{\milli\electronvolt}$ at room temperature and the calculated binding energy curves (\autoref{fig:geometry-dependence}c), we should expect the molecule to move much less than \SI{1}{\angstrom} along the distance axis, but in reality the molecule is free to move laterally as well as rotate.
Sampling the landscape of \gls{hc} transfer in the full space of molecular movement is a high-dimensional problem that can be addressed in future work.
For the purpose of a quantitative comparison to experimental realizations, it is crucial to consider that while metal states are accurately described with our level of theory \cite{KuiOjaEnk10}, molecular states are not.
We should thus expect transfer maxima to occur for different geometrical configurations, due to slightly shifted (hybridized) molecular orbitals; however, our general conclusions are still valid.

The importance of ground state hybridization for \gls{hc} transfer implies that theoretical modeling should not be restricted to considering bare metal surfaces, without taking interactions with molecules into account.
The distance-dependent hybridization of molecular orbitals should be explicitly taken into account for meaningful predictions.
Our results suggest that since already the ground state of the hybridized system is a good descriptor for prediction of \gls{hc} generation on molecules, rapid screening of candidates of good systems can be performed, without conducting expensive real-time simulations.

We close by commenting on the possibilities for tuning \gls{hc} transfer suggested by the results of this study.
\Gls{hc} devices can be designed by tuning the resonance condition to achieve a desired purpose; handles for tuning to a certain molecular orbital are the \gls{np} \gls{dos}, surface substitutions that affect the hybridization of the orbital, and the frequency of the incoming light.
As we have demonstrated, the tuning of the latter influences the absorption cross section so that there is a trade-off between high fraction of carriers transferred and high amount of carriers generated in total.
It is possible, however, to shift the absorption maximum with a rather small impact on level alignment, by modifying the \gls{np} shape and size, so that there is a maximum in both absorption and transfer.
In this way one ought to be able to maximize \gls{hc} transfer.
Furthermore, the sharp \gls{lsp} resonances of Ag \glspl{np} could possibly be utilized in the design of highly selective catalysts that work with broadband (solar) light; if the \gls{np} \gls{pdos} consists of one particularly strong peak, and that peak is resonant to one specific molecular orbital with the  frequency of the \gls{lsp} resonance, then transfer to that specific orbital will be preferred over transfer to other orbitals.

\section*{Data Availability}
The data generated in this study are openly available via Zenodo at \url{https://doi.org/10.5281/zenodo.6524101}, Ref.~\onlinecite{FojRosErh22Data}.

\section*{Software used}
The VASP\cite{KreHaf93, KreFur96, KreFur96a, KreJou99} suite with the \gls{paw}\cite{Blo94} method and the vdW-df-cx\cite{BerHyl14, KliBowMic09, KliBowMic11, RomSol09} \gls{xc}-functional was used for the total energy calculations and structure relaxations.
The \gpaw{} package \cite{MorHanJac05, EnkRosMor10} with \gls{lcao} basis sets \cite{LarVanMor09} and the \gls{lcao}-\gls{rttddft} implementation \cite{KuiSakRos15} was used for the \gls{rttddft} calculations.
The \textsc{ase} library \cite{LarMorBlo17} was used for constructing and manipulating atomic structures.
The NumPy \cite{HarMilvan20}, SciPy \cite{VirGomOli20} and Matplotlib \cite{Hun07} Python packages and the VMD software \cite{HumDalSch96, Sto98} were used for processing and plotting data.
The Snakemake \cite{MolJabLet21} package was used for managing the calculation workflow.

\section*{Acknowledgments}
We gratefully acknowledge helpful discussions with Mikael Kuisma.
This research has been funded by the Knut and Alice Wallenberg Foundation (2015.0055, 2019.0140; J.F., P.E.), the Swedish Foundation for Strategic Research Materials framework (RMA15-0052; J.F., P.E.), the Swedish Research Council (2015-04153, 2020-04935; J.F., P.E.), the European Union's Horizon 2020 research and innovation programme under the Marie Sk{\l}odowska-Curie grant agreement No~838996 (T.P.R.), and by the Academy of Finland under grant No~332429 (T.P.R.).
The computations were enabled by resources provided by the Swedish National Infrastructure for Computing (SNIC) at NSC, C3SE and PDC partially funded by the Swedish Research Council through grant agreement no. 2018-05973 as well as by the CSC -- IT Center for Science, Finland, and by the Aalto Science-IT project, Aalto University School of Science.

\appendix
\section{Methodology}
\label{sec:methods}

In the limit of weak perturbation, the system response can be assumed to be linear.
This means that the Fourier transform of the \emph{response} at one particular frequency depends only on the Fourier transform of the \emph{perturbation} at the same frequency.
The response to one time-dependent perturbation can be retrieved knowing the response to another perturbation.
We use this property in \gls{rttddft} to compute optical spectra (\autoref{subsec:rttddft}) and model carrier generation in arbitrary electric fields (\autoref{subsec:hc-generation}).
The geometrical configurations of the systems we studied are presented in \autoref{subsec:geometry-details} and computational details including \gls{dft} and \gls{rttddft} parameter values in \autoref{subsec:computational-details}.
For simplicity we adopt Hartree atomic units in this section.

\subsection{Plasmonic response in the \deltakick{} technique}
\label{subsec:rttddft}

In the \deltakick{} technique \cite{YabBer96} the time-dependent external potential is set to a kick in the $z$-direction of strength $K_z$
\begin{align}
    v_\text{ext}(\vec{r}, t) = K_z z \delta(t),
\end{align}
and the system is allowed to evolve in time.

We define the induced density due to the \deltakick{} $\delta n^\text{kick}(\vec{r}, t) = n^\text{kick}(\vec{r}, t) - n(\vec{r}, 0)$ and the induced dipole moment in the $z$-direction as
\begin{align}
    \delta \mu^\text{kick}_z(t)
    = \mu^\text{kick}_z(t) -  \mu_z(0)
    = \int \dd{\vec{r}} \delta n^\text{kick}(\vec{r}, t) z.
\end{align}

The optical spectrum in the $z$ direction can then be expressed via the dipole strength function
\begin{align}
    S_z(\omega) = -\frac{2\omega}{\pi} \Im\left[\frac{\mu_z(\omega)}{K_z}\right],
\end{align}
where $\mu_z(\omega)$ is the Fourier transform of $\mu_z(t)$.
% The dipole strength function is normalized so that it integrates to the number of electrons $N_e$ in the system
% \begin{align}
%     \int_0^\infty S_z(\omega) \dd{\omega} = N_e.
% \end{align}

\subsection{Carrier generation in the \deltakick{} technique}
\label{subsec:hc-generation}

In \gls{ks}-\gls{rttddft} the \gls{ks} density operator
\begin{align}
    \rho(t) = \sum_k \ket{\phi_k(t)} f_k \bra{\phi_k(t)}
\end{align}
contains all information about the system.
Here, $\phi_k(t)$ are the \gls{ks} wave functions at time $t$ and $f_k$ their ground state occupation numbers.
In practice, the operator is expressed as the \gls{ks} density matrix in the basis of ground state wave functions $\ket{\phi^{(0)}_k} = \ket{\phi_k(0)}$
\begin{align}
    \rho_{nn'}(t)
    &= \bra{\phi^{(0)}_n}\rho(t)\ket{\phi^{(0)}_{n'}} \\
    &= \sum_k \bra{\phi^{(0)}_n}\ket{\phi_k(t)} f_k \bra{\phi_k(t)}\ket{\phi^{(0)}_{n'}}.
\end{align}

In linear response we can express the \gls{ks} density matrix $\rho_{nn'}(t)$ corresponding to an arbitrary time-dependent field $\vec{\mathcal{E}}(t)$ in terms of the \gls{ks} density matrix $\rho^\text{kick}_{nn'}(t)$ due to the \deltakick{}
\begin{align}
    \rho_{nn'}(\omega) = \frac{1}{K_z} \rho^\text{kick}_{nn'}(\omega) \mathcal{E}_z(\omega).
    \label{eq:rhofromrhokick}
\end{align}
Here, functions of $\omega$ are Fourier transforms of the corresponding time-dependent quantities, and we have constrained ourselves to fields where the $z$-component $\mathcal{E}_z$ is the only non-zero component of $\vec{\mathcal{E}}$.
By the convolution theorem \autoref{eq:rhofromrhokick} is equivalent to
\begin{align}
    \rho_{nn'}(t) = \frac{1}{K_z} \int_0^\infty \dd{\tau} \rho^\text{kick}_{nn'}(\tau) \mathcal{E}_z(t-\tau).
\end{align}

In practice, we compute the Fourier transform of the \gls{ks} density matrix during the time propagation with the \deltakick{} and obtain the response to other fields as a post-processing step, using \autoref{eq:rhofromrhokick} and inverse Fourier transformation.

% While the $\delta\rho_{ia}(t)$ obtained from \gls{rttddft} calculations is a first-order quantity in perturbation we add second order corrections in the following expressions.
To compute the distributions of generated carriers we follow the method of Ref.~\citenum{RosErhKui20}, where expressions that include second-order corrections to the density matrix are derived.
For notational convenience (see Supplementary Note 2 of Ref.~\citenum{RosErhKui20}), we introduce the notation
\begin{align}
    q_{ia}(t) &= 2\Re \delta\rho_{ia}(t) \big/ \sqrt{2(f_i - f_a)} \\
    p_{ia}(t) &= -2\Im \delta\rho_{ia}(t) \big/ \sqrt{2(f_i - f_a)},
\end{align}
which is well-defined for matrix elements corresponding to occupied ($i$) -- unoccupied ($a$) pairs, $f_i > f_a$.
Only these occupied--unoccupied pairs are needed to construct the observables of interest.

The probability that a hole (electron) has been generated at time $t$ in state $i$ ($a$) is
\begin{align}
    \label{eq:hole-generation}
    P_i^\mathrm{h}(t) &= \sum_a^{f_i > f_a} \frac{1}{2} \left[q_{ia}^2(t) + p_{ia}^2(t)\right]\\
    \label{eq:electron-generation}
    P_a^\mathrm{e}(t) &= \sum_i^{f_i > f_a} \frac{1}{2} \left[q_{ia}^2(t) + p_{ia}^2(t)\right].
\end{align}
The sum of generated carriers is the same for both types
\begin{align}
    N_\mathrm{carriers}(t) = \sum_i P_i^\mathrm{h}(t) = \sum_a P_a^\mathrm{e}(t).
\end{align}

\begin{widetext}
    These probabilities are alternatively expressed as spatial probability densities (note that we drop the explicit $t$ dependence in $q$ and $p$ for brevity)
    \begin{align}
        \label{eq:spatial-hole-generation}
        P_\mathrm{h}(\vec{r}, t) &= \frac{1}{2} \sum_{aii'}^{f_i > f_a, f_{i'} > f_a}
                           (q_{ia}q_{i'a} + p_{i'a}p_{ia}) \psi_i^{(0)}(\vec{r})\psi_{i'}^{(0)}(\vec{r})\\
        \label{eq:spatial-electron-generation}
        P_\mathrm{e}(\vec{r}, t) &= \frac{1}{2} \sum_{iaa'}^{f_i > f_a, f_i > f_{a'}}
                           (q_{ia}q_{ia'} + p_{ia}p_{ia'}) \psi_a^{(0)}(\vec{r})\psi_{a'}^{(0)}(\vec{r}).
    \end{align}
    Note that integrating \autoref{eq:spatial-hole-generation} and \autoref{eq:spatial-electron-generation} over the entire space yields the number density of carriers $\int \dd{\vec{r}} P^\mathrm{e/h}(\vec{r}, t) = N_\mathrm{carriers}(t)$.
    Further, we construct spatio-energetic probability density contributions
    \begin{align}
        \label{eq:spatioenergetic-hole-generation}
        P_\mathrm{h}(\varepsilon, \vec{r}, t) &= \frac{1}{2} \sum_{aii'}^{f_i > f_a, f_{i'} > f_a}
                                        (q_{ia}q_{i'a} + p_{i'a}p_{ia}) \psi_i^{(0)}(\vec{r})\psi_{i'}^{(0)}(\vec{r})
                                        \delta(\varepsilon - \varepsilon_i) \\
        \label{eq:spatioenergetic-electron-generation}
        P_\mathrm{e}(\varepsilon, \vec{r}, t) &= \frac{1}{2} \sum_{iaa'}^{f_i > f_a, f_i > f_{a'}}
                                        (q_{ia}q_{ia'} + p_{ia}p_{ia'}) \psi_a^{(0)}(\vec{r})\psi_{a'}^{(0)}(\vec{r})
                                        \delta(\varepsilon - \varepsilon_a).
    \end{align}
    The $\delta$-functions in energy are in practice approximated by a Gaussian $(2\pi\sigma^2)^{-1/2} \exp(-\varepsilon^2/2\sigma^2)$ with width $\sigma=\SI{0.07}{\electronvolt}$.
    There is no unique definition for the simultaneous decomposition in time and energy but we obtain consistent results (\autoref{sfig:summation-comparison}) with different variants of the former expressions.
    Note that in contrast to Ref.~\citenum{RosErhKui20} we include the summation of non-degenerate $ii'$ and $aa'$ which has significant implications in the \gls{np}+molecule system (\autoref{sfig:summation-comparison}).
\end{widetext}

In this work, we calculate the fraction of generated carriers in particular regions, i.e., the Voronoi region of the molecule or the Voronoi region of a particular atomic site
\begin{align}
    \label{eq:region-hole-generation}
    P^\text{h,region}(t) &= \int_\text{region}\mathrm{d}\vec{r} P_\mathrm{h}(\vec{r}, t) \big/ N_\mathrm{carriers}(t) \\
    \label{eq:region-electron-generation}
    P^\text{e,region}(t) &= \int_\text{region}\mathrm{d}\vec{r} P_\mathrm{e}(\vec{r}, t) \big/ N_\mathrm{carriers}(t).
\end{align}
Similarly, we calculate the energetic distribution in a particular region
\begin{align}
    \label{eq:region-energetic-hole-generation}
    P^\text{h,region}(\varepsilon, t) &= \int_\text{region}\mathrm{d}\vec{r} P_\mathrm{h}(\varepsilon, \vec{r}, t) \\
    \label{eq:region-energetic-electron-generation}
    P^\text{e,region}(\varepsilon, t) &= \int_\text{region}\mathrm{d}\vec{r} P_\mathrm{e}(\varepsilon, \vec{r}, t).
\end{align}

\subsection{Geometry of atomic structures}
\label{subsec:geometry-details}

In this work we study the \gls{rto}-shaped \ce{Ag201}, \ce{Au201}, and \ce{Cu201} \glspl{np}.
The \gls{rto} shape is consistent with an unstrained \gls{fcc} lattice that has been truncated so that it has eight \{111\} surfaces and six \{100\} surfaces.
We identify four molecular adsorption sites on the \gls{np}.
On the (111) surface we consider the center on-top site and its nearest \gls{fcc} site (\autoref{fig:geometry-dependence}).
We also consider the hollow site closest to the center of the (100) face, and the corner site between a \{100\} and two \{111\} surfaces.
In each site the immediate environment of the molecule is different; it has one nearest-neighbor metal atom for the (111) on-top and corner sites, three for the (111) \gls{fcc} site and four for the (100) hollow site.

We place a CO molecule (C facing the metal) near each site and find minimum energy configurations by allowing all metal and molecule atoms to relax.
In the relaxed configurations the bond length and \gls{np}-molecule distance are thus different for each site.

We study the distance dependence by rigidly shifting the molecule from the adsorbed configuration, i.e., without changing the CO bond length or the positions of the metal atoms.
The shift is performed along a line perpendicular to the respective surface for the (111) and (100) sites and along the line through the opposing corner site for the corner site.
We note that at long separations such rigidly shifted configurations are relatively high in energy.
Specifically, at long distances the minimum energy configuration would be a molecule with the bond length of bare CO and a fully relaxed bare \gls{np}.
In general, finding such minimum energy configurations under a distance constraint is a highly multidimensional problem.
We therefore limit our discussion to rigid translations of the molecule, without compromising our conclusions (\autoref{sfig:level-alignment}).

\subsection{Computational details}
\label{subsec:computational-details}

The VASP \cite{KreHaf93, KreFur96, KreFur96a} suite was used for all structure relaxations and total energy calculations.
Ground state energies were computed using a plane-wave basis set, the \gls{paw} \cite{Blo94, KreJou99} method, and the vdW-df-cx \cite{DioRydSch04, BerHyl14, KliBowMic09, RomSol09} \gls{xc}-functional.
The plane wave cutoff \SI{500}{\electronvolt} was used.
%In each electronic ground state computation the stopping criterion of the self-consistency loop was that subsequent energies should differ by less than \SI{1e-5}{\electronvolt}.
We used a Gaussian occupation number smearing scheme with the parameter \SI{0.1}{\electronvolt}.

Structure relaxations were performed using the conjugate gradient relaxation method implemented in VASP.
Relaxation was stopped when the maximal force on any atom fell below \SI{0.015}{\electronvolt\per\angstrom}.
Relaxations were performed for each \gls{np} separately to obtain relaxed bare \gls{np} structures.
Additionally for each \gls{np} and each adsorption site the \gls{np}+molecule system was relaxed, with all atoms allowed to move.

The open-source \gpaw{}\cite{MorHanJac05, EnkRosMor10} code package was used for all calculations of plasmonic response and carrier generation.
\Gls{ksdft} ground state calculations were performed within the \gls{paw} \cite{Blo94} formalism using \gls{lcao} basis sets \cite{LarVanMor09}; the \gls{dzp} basis set was used for C and O, and the \emph{pvalence} \cite{KuiSakRos15} basis set, which is optimized to represent bound unoccupied states, for Ag, Au and Cu.
The \gls{gllbsc} \cite{GriLeeLen95, KuiOjaEnk10} \gls{xc} functional was used.
A simulation cell of \SI{32x32x38.4}{\angstrom} was used to represent wave functions, \gls{xc}, and Coulomb potentials, with a grid spacing of \SI{0.2}{\angstrom} for wave functions and \SI{0.1}{\angstrom} for potentials.
The Coulomb potential was represented in numerical form on the grid, with an additional analytic moment correction \cite{CasRubSto03} centered at the \gls{np}.
Fermi-Dirac occupation number smearing with width \SI{0.05}{\electronvolt} was used for all calculations, except for the Cu corner site for distances $\geq$\SI{2.7}{\angstrom} and for Au (111) on-top distances between \SI{3.1}{} and \SI{3.7}{\angstrom}.
The effect of different occupation number smearing is negligible for our results (\autoref{sfig:fdsmearing}).
The self-consistent loop was stopped when the integral of the difference between two subsequent densities was less than \SI{1e-8}{}.
Pulay \cite{Pul80}-mixing was used to accelerate the ground state convergence.

The \gls{lcao}-\gls{rttddft} implementation \cite{KuiSakRos15} in \gpaw{} was used for the \gls{rttddft} calculations.
A \deltakick{} strength of $K_z = \SI{0.51e-4}{\milli\volt\per\angstrom}$ was used.
The time propagation was done in steps of \SI{10}{\atto\second} for a total length of \SI{30}{\femto\second} using the adiabatic \gls{gllbsc} kernel.
We computed carrier generation for an external electric field corresponding to an ultra-short Gaussian laser pulse
\begin{align}
    \label{eq:laser}
    \mathcal{E}_z(t) = \mathcal{E}_0 \cos(\omega_0 t) \exp(- (t-t_0)^2/\tau_0^2)
\end{align}
of frequency $\omega$, strength $\mathcal{E}_0 =  \SI{51}{\micro\volt\per\angstrom}$, peak time $t_0 = \SI{10}{\femto\second}$, and duration $\tau_0 = \SI{3.0}{\femto\second}$.

We computed the total \gls{dos} as
\begin{align}
    \sum_k \delta(\varepsilon - \varepsilon_k)
\end{align}
and the molecular \gls{pdos} as
\begin{align}
    \sum_k \delta(\varepsilon - \varepsilon_k) \int_\text{molecule} \left|\phi^{(0)}_k(\vec{r})\right|^2 \dd{\vec{r}},
\end{align}
where $\varepsilon_k$ and $\phi^{(0)}_k(\vec{r})$ are the \gls{ks} eigenvalues and wave functions.
For visualization, the $\delta$-functions in energy were replaced by a Gaussian $(2\pi\sigma^2)^{-1/2} \exp(-\varepsilon^2/2\sigma^2)$ with width $\sigma=\SI{0.05}{\electronvolt}$.

\bibliography{main.bib}

\end{document}